\documentclass[twocolumn,pra,aps]{revtex4-1}
\usepackage[latin1]{inputenc}
\usepackage{amsmath}
\usepackage{amsfonts}
\usepackage{amssymb}
\usepackage{graphicx}
\usepackage{textcomp}
\usepackage{color}
\begin{document}
\title{Organization of the Hilbert Space for Exact Diagonalization of Hubbard Model}
\author{Medha Sharma}
\email{medhajamia@gmail.com}
\author{M.A.H. Ahsan}
\email{mahsan@jmi.ac.in}
\affiliation{Department of Physics, Jamia Millia Islamia, New Delhi 110025, India}

\begin{abstract}
We present an alternative scheme to the widely used method
of representing the basis of one-band Hubbard model through
the relation $I=I_{\uparrow}+2^{M}I_{\downarrow}$ given by H. Q. Lin and J. E. Gubernatis [Comput. Phys. 7, 400 (1993)], where
$I_{\uparrow}$, $I_{\downarrow}$ and $I$ are the integer equivalents of
binary representations of occupation patterns of spin up, spin
down and both spin up and spin down electrons respectively, with 
$M$ being the number of sites. We compute and store only $I_{\uparrow}$ or
$I_{\downarrow}$ at a time to generate the full Hamiltonian
matrix. The non-diagonal part of the Hamiltonian matrix given as
${\cal{I}}_{\downarrow}\otimes{\bf{H}_{\uparrow}} \oplus {\bf{H}_{\downarrow}}\otimes{\cal{I}}_{\uparrow}$
is generated using a bottom-up approach by computing the small matrices ${\bf{H}_{\uparrow}}$(spin up hopping Hamiltonian)
and ${\bf{H}_{\downarrow}}$(spin down hopping Hamiltonian)
and then forming the tensor product with respective identity matrices ${\cal{I}}_{\downarrow}$ and ${\cal{I}}_{\uparrow}$,
thereby saving significant computation time and memory. We find that the total CPU time to generate the non-diagonal part of the Hamiltonian
matrix using the new one spin configuration basis scheme is reduced by about an order of magnitude as compared to the two
spin configuration basis scheme.
The present scheme is shown to be inherently parallelizable. Its application to translationally invariant systems, computation of Green's functions and in impurity solver part of DMFT procedure is discussed and its extention to other models is also pointed out.
\end{abstract}

\maketitle
keywords:Hubbard model, Exact diagonalization

\section{Introduction}

Hubbard model\cite{jh1963} was introduced as an approximate model for
electron-electron interaction in narrow energy band systems and
serves as the simplest model that captures the
essence of strongly-correlated electrons in solids.
The one-band Hubbard model is
\begin{equation}
H= -t \sum_{<i,j>,\sigma}(c_{i \sigma}^{\dagger}c_{j \sigma}+
c_{j \sigma}^{\dagger}c_{i \sigma})+ U \sum_{i} n_{i\uparrow}n_{i\downarrow}\label{count1},
\end{equation}
where $c_{i \sigma}^{\dagger},c_{i \sigma}$ and $n_{i \sigma}=c_{i
\sigma}^{\dagger}c_{i \sigma} $ are fermion creation, annhilation and
number operators respectively at site $i$ with spin $\sigma (\uparrow or
\downarrow)$. The angular bracket $<i,j>$ denotes the sum over nearest neighbors.
The first term of the Hamiltonian describes the kinetic energy of the
itinerant electrons with $t$ being the hopping amplitude and the second term contains the on-site 
Coulomb repulsion $U$.

An exact solution of Hubbard model in one dimension was given by
Lieb and Wu\cite{lw1968} using Bethe Ansatz. The infinite lattice
coordination limit introduced by Metzner and Vollhardt\cite{mv1989}
forms the basis for the Dynamical Mean Field Theory(DMFT)
that maps the Hubbard model onto an Anderson impurity model\cite{ag1992}.
Except for the above two extreme cases, numerical methods are required to solve the 
Hamiltonian (\ref{count1}) and turns out to be quite formidable for any system of interest\cite{ed1994}.
DMFT can be applied as an approximation scheme directly to three-dimensional lattice problems.

Exact Diagonalization(ED)\cite{ed1994,sr1989,jd1990,pj2011} is an important technique for studying quantum
many-body systems. It provides information about many-body correlations
by giving an exact solution of the model albeit on finite systems.
It is a memory expensive technique, apart from being CPU intensive.
ED solver is limited to small clusters because of large 
memory required for an exponentially growing Hilbert space.
It also provides flexibility on parameters tuning. When used as an impurity 
solver in DMFT scheme, ED introduces parametrization of the effective bath 
which makes it superior to Quantum Monte Carlo technique\cite{ck1994}.
Unlike Quantum Monte Carlo simulations at low temperatures, ED
suffers no fermionic sign problem. With proper finite size scaling, 
it helps to gain insight into the many-body system in thermodynamic
limit. ED is also advantageous in providing real-frequency 
information and serves as a check on approximate methods.

High temperature cuprate superconductors are described by three-band
Hubbard model\cite{vj1987}, but it has been argued that the 
essential physics can be captured by a one-band Hubbard model\cite{pw1987,zr1988}.
Physics of pseudogap in cuprate superconductors has be studied using extended dynamical mean field theory with ED as impurity solver\cite{al2009,nl2010}. 
Pariser-Parr-Popple model\cite{rr1953,ja1953} on which lot of ED work is done in quantum chemistry and molecular systems is a natural extention to Hubbard model\cite{ja2010}.

The scheme to generate the basis states of one-band Hubbard model using both up and down spin was given by
H. Q. Lin and J. E. Gubernatis\cite{lg1993} about two decades ago. Since
then researchers have been using\cite{ym2007,sh2012,ds2010,aj2008,aw2008}
the two spin configuration basis.
Since in the Hubbard model, up and down spins do not mix with
each other as no term in the Hamiltonian changes an up spin to a down
spin and vice versa, we treat both spin bases in a many-body basis state
separately while applying the Hamiltonian to them and present a new scheme that requires only one
spin configuration basis at a time to generate the Hamiltonian matrix leading to significant gain in computation time and memory.

For basis construction of Hubbard model, the most commonly used symmetries are particle number conservation, $S_{z}$ conservation and translational invariance\cite{aw2010}.

The contents of this paper are organised as follows. In section II, we describe the generation of basis states in our scheme.
Section III presents the generation of the Hamiltonian matrix.
In Section IV, we demonstrate the usefullness of our scheme in diagonalization of the Hamiltonian matrix.
Section V presents some test runs to compare two-spin and one-spin basis schemes.
In Section VI, we demonstrate the use of our scheme on translationally symmetric systems.
Section VII describes the computation of one-particle Green's functions in our scheme. In Section VIII, we discuss the application of our scheme in DMFT. In Section IX we extend our scheme to other models.
Finally in section X, we discuss the advantages of our scheme over the widely used two spin configuration basis scheme.

\section{Generation of basis states}

The number operator:
\begin{equation}
N=\sum_{i}(n_{i\uparrow}+n_{i\downarrow})=N_{\uparrow}+N_{\downarrow} 
\end{equation}
and the z-projection of the total spin:
\begin{equation}
S^{z}=\frac{1}{2}\sum_{i}(n_{i\uparrow}-n_{i\downarrow})=\frac{1}{2}(N_{\uparrow}-N_{\downarrow})
\end{equation}
both commute with the Hamiltonian (\ref{count1}). For basis construction we
use both these symmetries which is equivalent to conservation of
total number of spin up$(N_{\uparrow}=\frac{1}{2}N+S^{z})$ and total number of spin
down$(N_{\downarrow}=\frac{1}{2}N-S^{z})$ electrons. We perform diagonalization in a
sector$(N_{\uparrow},N_{\downarrow})$ of total Hilbert space with
fixed number of spin up electrons and of spin down
electrons. Hilbert space for a sector$(N_{\uparrow},N_{\downarrow})$
can be constructed by forming the tensor product:
\begin{equation}
{\cal{V}}^{(N_{\uparrow},N_{\downarrow})}={\cal{V}}^{N_{\uparrow}}\otimes{\cal{V}}^{N_{\downarrow}} .
\end{equation}
The basis states spanning ${\cal{V}}^{(N_{\uparrow},N_{\downarrow})}$ map uniquely onto an integer $I$ defined by\cite{lg1993}
\begin{equation}
I=\sum_{i=1}^{M}[n_{\uparrow}(i)2^{(i-1)}+n_{\downarrow}(i)2^{(M+i-1)}],
\end{equation}
where $n_{\uparrow}(i)$ and $n_{\downarrow}(i)$ are the occupancies of site
$i$ for up spin and down spin respectively; M being the total number of sites.
The basis state of ${\cal{V}}^{(N_{\uparrow},N_{\downarrow})}$ is written as:
\begin{equation}
I=I_{\uparrow}+2^{M}I_{\downarrow},
\end{equation} 
where $I_{\uparrow}=\sum_{i=1}^{M}n_{\uparrow}(i)2^{(i-1)}$ is the spin up basis state of ${\cal{V}}^{N_{\uparrow}}$ and $I_{\downarrow}=\sum_{i=1}^{M}n_{\downarrow}(i)2^{(i-1)}$ is the spin down basis state of ${\cal{V}}^{N_{\uparrow}}$ .

The bits of integer $I$ represent a specific basis state:
\begin{eqnarray}
&&\left|n_{\uparrow}(1),n_{\uparrow}(2),...,n_{\uparrow}(M)\right\rangle\left|n_{\downarrow}(1),n_{\downarrow}(2),...,n_{\downarrow}(M)\right\rangle\nonumber\\
\noindent&=&(c_{1\uparrow}^{\dagger})^{n_{1\uparrow}}...(c_{M\uparrow}^{\dagger})^{n_{M\uparrow}}(c_{1\downarrow}^{\dagger})^{n_{1\downarrow}}...(c_{M\downarrow}^{\dagger})^{n_{M\downarrow}}\left|0\right\rangle.
\end{eqnarray}

Like the two-table method of Lin\cite{hq1990}, we store $I_{\uparrow}^{,}s$ and $I_{\downarrow}^{,}s$ separately.
In our scheme, we assign serially ordered indices starting from 1 to spin up basis states spanning ${\cal{V}}^{N_{\uparrow}}$ and to spin down 
basis states spanning ${\cal{V}}^{N_{\downarrow}}$ and use a relation:
\begin{equation}
Index=(Index_{\downarrow}-1)count_{\uparrow}+Index_{\uparrow}\label{count2},
\end{equation}
where $Index$=index of basis state ($I$) of ${\cal{V}}^{(N_{\uparrow},N_{\downarrow})}$,
$Index_{\uparrow}$=index of spin up basis state ($I_{\uparrow}$) of ${\cal{V}}^{N_{\uparrow}}$, $Index_{\downarrow}$=index
of spin down basis state ($I_{\downarrow}$) of ${\cal{V}}^{N_{\downarrow}}$, $count_{\uparrow}$=total number of basis states of spin
up configuration spanning ${\cal{V}}^{N_{\uparrow}}$. The algebraic relation (\ref{count2}) give the index of $I$ of ${\cal{V}}^{(N_{\uparrow},N_{\downarrow})}$ in terms of index of $I_{\uparrow}$ of ${\cal{V}}^{N_{\uparrow}}$ and of $I_{\downarrow}$ of ${\cal{V}}^{N_{\downarrow}}$ respectively.

\begin{table}
\begin{tabular}{|c|c|c|c|c|c|c|}
	\hline
$Up$  & $Basis_{\uparrow}$ & $Index_{\uparrow}$ & $Down$  & $Basis_{\downarrow}$ & $Index_{\downarrow}$ & $Index$  \\
 & & & &  & & \\
	\hline
011 & 3 & 1 & 001 & 1 & 1 & 1 \\
101 & 5 & 2 & 001 & 1 & 1 & 2 \\
110 & 6 & 3 & 001 & 1 & 1 & 3 \\
011 & 3 & 1 & 010 & 2 & 2 & 4 \\
101 & 5 & 2 & 010 & 2 & 2 & 5 \\
110 & 6 & 3 & 010 & 2 & 2 & 6 \\
011 & 3 & 1 & 100 & 4 & 3 & 7 \\
101 & 5 & 2 & 100 & 4 & 3 & 8 \\
110 & 6 & 3 & 100 & 4 & 3 & 9 \\
\hline
\end{tabular}
\caption{Spin up and down configurations, their bases, their indices
and indices of the complete states representing both spin configurations
for $M=3$ in a $sector(N_{\uparrow}=2,N_{\downarrow}=1)$}
\end{table}
For example, from table I, the spin up basis state $\left|110\right\rangle$ of ${\cal{V}}^{N_{\uparrow}}$ having $index_{\uparrow}$=3 and spin down 
basis state $\left|010\right\rangle$ of ${\cal{V}}^{N_{\downarrow}}$ having $index_{\downarrow}$=2 will result in a basis state 
$\left|110\right\rangle\left|010\right\rangle$ of ${\cal{V}}^{(N_{\uparrow},N_{\downarrow})}$ with an index=(2-1)3+3=6.

All the basis states spanning ${\cal{V}}^{N_{\sigma}}$ where $(\sigma=\uparrow or \downarrow)$ and their respective 
indices are generated using Algorithm I, given
in Appendix \ref{algo1} which assumes the existence of a bit fuction
\textbf{bittest}$(i,j)$ that returns true if bit in position $j$ of $i$ is 
1, else false; $i$ and $j$ being integers. Most of high level programming languages such as fortran 90 and $C^{++}$ have intrinsic
functions for bitwise operations on integers.

Total number of basis states spanning ${\cal{V}}^{N_{\sigma}}$ where $(\sigma=\uparrow or
\downarrow)$ , i.e., $count_{\sigma}$ is computed using the number of
ways to distribute $N_{\sigma}$ electrons among $M$ sites.
\begin {equation}
count_{\sigma}= {^{M}C}_{N_{\sigma}}=\frac{M!}{N_{\sigma}!(M-N_{\sigma})!} \label{count3}
\end{equation}

The dimensionality of Hilbert space of a given sector $(N_{\uparrow},N_{\downarrow})$ is
\begin{equation} 
dimension{\cal{V}}^{(N_{\uparrow},N_{\downarrow})}=count_{\uparrow}count_{\downarrow}
\end{equation} 
but we are not required to generate 
$I^{,}s$ spanning the full Hilbert space ${\cal{V}}^{(N_{\uparrow},N_{\downarrow})}$. We work only with  
$I_{\uparrow}^{,}s$ spanning ${\cal{V}}^{N_{\uparrow}}$ or $I_{\downarrow}^{,}s$ spanning ${\cal{V}}^{N_{\downarrow}}$ at a time.

\section{Generation of the Hamiltonian matrix}

When the Hamiltonian is applied to each of the basis states, the
Hamiltonian matrix is generated.

\subsection{Non-diagonal part of the Hamiltonian matrix}

The non-diagonal part of the Hamiltonian
matrix is due to the effect of hopping terms
$c_{i\sigma}^{\dagger}c_{j\sigma}+c_{j\sigma}^{\dagger}c_{i\sigma}$
that move an electron from site $i$ to $j$ or from $j$ to $i$.

We generate all the spin $\sigma$ basis states spanning ${\cal{V}}^{N_{\sigma}}$ (using
Algorithm I, given in Appendix \ref{algo1}) and compute the total number of spin $\overline{\sigma}$
basis spanning ${\cal{V}}^{N_{\overline{\sigma}}}$, i.e., $count_{\overline{\sigma}}$ using Eq.(\ref{count3}), where if $\sigma=\uparrow(\downarrow)$ then $\overline{\sigma}=\downarrow(\uparrow)$.

Let the action of of a spin $\sigma$ hopping term change a state $basis_{\sigma}(p)$ to
$basis_{\sigma}(l)$,

\begin {equation}
-t(c_{i\sigma}^{\dagger}c_{j\sigma}+c_{j\sigma}^{\dagger}c_{i\sigma})basis_{\sigma}(p)=-t(esign)basis_{\sigma}(l),
\end {equation}

where $esign$ takes care of the sign depending upon the number of occupied sites between the $i$ 
and $j$ sites, i.e., if an electron hops over an even 
number of electrons, $esign=+1$ and if it hops over an odd number of electrons, $esign=-1$.
$Index_{\sigma}$ $p$ is known and the $index_{\sigma}$ $l$ can be found either by storing the indices of the basis states
in a seperate array while generating the basis states as in Algorithm I or by using a binary search.

All the elements of the Hamiltonian matrix due to this particular hopping between 
these two spin $\sigma$ basis states can be computed via a simple loop, for spin $\uparrow$:

\noindent\textbf{for} $k=1:count_{\downarrow}$\\
\indent\indent $ r=(k-1)count_{\uparrow}+p $\\
\indent\indent $ s=(k-1)count_{\uparrow}+l $\\
\indent\indent $ matrix(r,s)=-t(esign)$\\
\textbf{end}\\

and for spin $\downarrow$:

\noindent\textbf{for} $k=1:count_{\uparrow}$\\
\indent\indent $r=(p-1)count_{\uparrow}+k$\\
\indent\indent $s=(l-1)count_{\uparrow}+k$\\
\indent\indent $matrix(r,s)=-t(esign)$\\
\textbf{end},\\

where $matrix(r,s)$ is the (r,s)th element of the Hamiltonian matrix. All the matrix elements 
generated through this $\sigma=\uparrow or \downarrow$ loop are obtained by applying a spin $\sigma$ hopping term to a spin $\sigma$ basis 
state and enables us to get rid of the repetitive application of the Hamiltonian everytime to get a 
matrix element\cite{aw2008}.

The above procedure is repeated for each of the spin $\sigma$ hopping terms acting on each of the spin $\sigma$ 
basis states spanning ${\cal{V}}^{N_{\sigma}}$ to obtain all the matrix elements due to spin $\sigma$ hopping.

\subsection{Diagonal part of the Hamiltonian matrix}

The diagonal part of the Hamiltonian matrix is due to the onsite Coulomb
 interaction that counts the double occupancy of a site. The onsite
interaction term $U\sum_{i}n_{i\uparrow}n_{i\downarrow}$ acting on a
basis state gives the same basis state multiplied by the number of doubly
occupied sites times $U$.

For generation of diagonal matrix elements the basis states of only one 
configuration $(\sigma=\uparrow or \downarrow)$ spanning ${\cal{V}}^{N_{\sigma}}$ are required at a time.

All the diagonal elements of the Hamiltonian matrix can be computed using
Algorithm II, given in Appendix \ref{algo2} in which $matrix(r,r)$ is the
(r,r)th element of the Hamiltonian matrix and $U$ being the onsite interaction. In Algorithm II, if $(N_{\uparrow}+N_{\downarrow})\leqslant M$ then $point_{\uparrow}={^{M-1}C}_{N_{\uparrow}-1}
(point_{\downarrow}={^{M-1}C}_{N_{\downarrow}-1})$ is the total number of spin up(down) basis states of 
${\cal{V}}^{N_{\uparrow}}({\cal{V}}^{N_{\downarrow}})$ in which any one given site is occupied and if $(N_{\uparrow}+N_{\downarrow})>M$ then $point_{\uparrow}={^{M-1}C}_{N_{\uparrow}}
(point_{\downarrow}={^{M-1}C}_{N_{\downarrow}})$ is the total number of spin up(down) basis states of
${\cal{V}}^{N_{\uparrow}}({\cal{V}}^{N_{\downarrow}})$ in which any one given site is unoccupied.
In sectors where either $N_{\uparrow}$ or $N_{\downarrow}$ is equal to $M$, all the diagonal elements are equal to $U$ times $min(N_{\uparrow},N_{\downarrow})$.

\section{Diagonalization of the Hamiltonian matrix}

After generating the Hamiltonian matrix we diagonalize it to find the eigenvalues and eigenvectors. Owing to the spareness of the Hamiltonian matrix and the fact that we are interested in the eigenvalues and eigenvectors of the ground state and a few low-lying excited states only, we can use the Lanczos method\cite{ln1950} of diagonalizing large, sparse, symmetric matrices. For using the Lanczos algorithm, the matrix does not have to be constructed explicitly, since only its application to a vector is needed to compute the span of the Krylov subspace $ {K_{j}(H,q)=q,Hq,...,H^{j-1}q}$. The main computational step in the Lanczos Algorithm is the matrix-vector multiplication without having an explicit representation of the matrix. One way is to have some functional representation of the matrix taking its repeating patterns into account so that it can be applied to a vector and the other way is to compute the Hamiltonian matrix everytime as and when required. Our scheme will be useful in both the cases.

\subsection{Storage of the nonzero elements of Hamiltonian matrix for matrix-vector multiplication}

In each row of the sparse Hamiltonian matrix, there are very few
nonzero elements. For the one-band Hubbard model on an one-dimensional
ring of $M$ sites, considering only the nearest-neighbours hopping,
the Hamiltonian matrix in any given row will have at the most $2M$
nonzero off-diagonal elements; $M$ elements due to the hopping terms
of either spin configuration $\sigma(\uparrow or \downarrow)$. For an
Anderson impurity model on an $M$ site lattice in which the trasition is
possible between the impurity site and the bath constituted by all other
sites, each row of the Hamiltonian matrix will have a maximum of $2M-2$
nonzero off-diagonal elements; $M-1$ elements due to the hopping terms
of either spin configuration $\sigma(\uparrow or \downarrow)$. Let $F$
be the maximum number of non-zero off-diagonal matrix elements in any given row
of the Hamiltonian matrix. For an $R\times R$ matrix, there is an effective
$R\times(F+1)$ matrix, where $R=count_{\uparrow}count_{\downarrow}$.

In our scheme both spin up and spin down bases are treated seperately.

The diagonal elements of the Hamiltonian matrix corresponding to
\begin{equation}
H_{U}=U\sum_{i} n_{i\uparrow}n_{i\downarrow}
\end{equation} 
can be generated on-the-fly by storing the integer arrays $index_{U\uparrow}$ and $index_{U\downarrow}$ of maximum dimension
$(M,\frac{1}{2}\frac{M!}{[(M/2)!]^{2}})$ for even $M$ and $(M,\frac{(M-1)!}{[(\frac{M-1}{2})!]^{2}})$ for odd $M$(Algorithm II, Appendix \ref{algo2}).

By applying the spin up hopping terms to each of the spin up basis 
states spanning ${\cal{V}}^{N_{\uparrow}}$, the following Hamiltonian:
\begin{equation}
H_{\uparrow}=-t \sum_{<i,j>}(c_{i \uparrow}^{\dagger}c_{j \uparrow} + c_{j \uparrow}^{\dagger}c_{i \uparrow})
\end{equation}
matrix can be generated having effective dimension $count_{\uparrow}\times(F/2)$.

Similarly by applying the spin down hopping terms to each of the spin
down basis states spanning ${\cal{V}}^{N_{\downarrow}}$, the following Hamiltonian:
\begin{equation}
H_{\downarrow}=-t\sum_{<i,j>}(c_{i \downarrow}^{\dagger}c_{j \downarrow} + c_{j \downarrow}^{\dagger}c_{i \downarrow})
\end{equation}
matrix can be generated having effective dimension $count_{\downarrow}\times(F/2)$.

The Hubbard Hamiltonian in matrix representation is mathematically given as:
\begin{equation}
\bf{H}={\cal{I}}_{\downarrow}\otimes{\bf{H}_{\uparrow}}  \oplus {\bf{H}_{\downarrow}}\otimes{\cal{I}}_{\uparrow}  \oplus \bf{H}_{U},
\end{equation}
where ${\cal{I}}_{\uparrow}({\cal{I}}_{\downarrow})$ is the identity operator for electrons with spin up(down).
Using our scheme we directly obtain the matrices $\bf{H_{\uparrow}}$ and $\bf{H_{\downarrow}}$ 
respectively.

Total number of non-zero matrix elements in $\bf{H_{\sigma}}$ where $(\sigma=\uparrow or \downarrow)$ 
is $F\times{^{M-2}C}_{N_{\sigma}-1}$, where $M>2$ and $0<N_{\sigma}<M$.
Thus number of non-zero matrix elements in $\bf{H}$ become
$F\times{^{M-2}C}_{N_{\uparrow}-1}\times count_{\downarrow}+F\times{^{M-2}C}_{N_{\downarrow}-1}\times count_{\uparrow}+count_{\uparrow}count_{\downarrow}$, where $M>2$, $0<N_{\uparrow}<M$ and $0<N_{\downarrow}<M$.

The storage of the non-diagonal part of {\bf{H}} using ${\bf{H_{\uparrow}}}$ and ${\bf{H_{\downarrow}}}$ is ideally suited for parallelization\cite{ym2007}.
We perform the matrix vector multiplication $q_{new}={\bf{H}}q_{old}$, where $q_{old}$ and $q_{new}$ are the vectors of dimension 
$count_{\uparrow}count_{\downarrow}$.
The vector product with the non-diagonal part of the Hamiltonian matrix due to spin up hoppping, i.e.
$q_{new}=({\cal{I}}_{\downarrow}\otimes{\bf{H_{\uparrow}}})q_{old}$ is reduced to the computation of $vecnew={\bf{H_{\uparrow}}}\times vecold$, where $vecold$ is the $count_{\uparrow}\times count_{\downarrow}$ matrix obtained from $q_{old}$ as follows:\\ 
$n=0$\\
\textbf{for} $j=1:count_{\downarrow}$\\
\indent\indent\textbf{for} $i=1:count_{\uparrow}$\\
\indent\indent\indent\indent$n=n+1$\\
\indent\indent\indent\indent$vecold(i,j)=q_{old}(n)$\\
\indent\indent\textbf{end}\\
\textbf{end}\\\\
and $vecnew$ is the $count_{\uparrow}\times count_{\downarrow}$ matrix that generates $q_{new}$ as follows:\\
$n=0$\\
\textbf{for} $j=1:count_{\downarrow}$\\
\indent\indent\textbf{for} $i=1:count_{\uparrow}$\\
\indent\indent\indent\indent$n=n+1$\\
\indent\indent\indent\indent$q_{new}(n)=vecnew(i,j)$\\
\indent\indent\textbf{end}\\
\textbf{end}.\\

$q_{new}=({\cal{I}}_{\downarrow}\otimes{\bf{H_{\uparrow}}})q_{old}$ can be parallelized using the Pseudocode given in Appendix \ref{pseudocode1}.

Similarly the vector product with the non diagonal part of the Hamiltonian matrix due to spin down hoppping, i.e.
$q_{new}=({\bf{H_{\downarrow}}}\otimes{\cal{I}}_{\uparrow})q_{old}$ is reduced to the computation of $vecnew=vecold\times{\bf{H_{\downarrow}}}^{T}$.
Matrices $\bf{H_{\uparrow}}$ and $\bf{H_{\downarrow}}$ can be stored on each node using sparse matrix format and matrix $vecold$ is distributed among all the nodes, thereby no inter-node communication
is required to carry out vector multiplication with non-diagonal part of {\bf{H}}. GPU (Graphics processing units) implementation using ${\bf{H_{\uparrow}}}$ and ${\bf{H_{\downarrow}}}$ has been described in Ref.14.

\subsection{Computation of non-zero matrix elements for matrix-vector multiplication}

In our scheme, to obtain the non-diagonal part of the Hamiltonian matrix we $\bf{compute}$ only the small matrix
$\bf{H_{\uparrow}}(\bf{H_{\downarrow}})$ and then form the tensor
product ${\cal{I}}_{\downarrow}\otimes{\bf{H}_{\uparrow}}({\bf{H}_{\downarrow}}\otimes{\cal{I}}_{\uparrow})$ through a nested loop.
In other words, for generation of the non-diagonal part of the Hamiltonian matrix due to spin $\sigma$ hopping terms, the operation of 
$H_{\sigma}$ acting on each of the $I^{,}s$  which is equal to $count_{\uparrow}count_{\downarrow}$ is split in our scheme
into the operation of $H_{\sigma}$ acting on each of the $I_{\sigma}^{,}s$ which is equal to $count_{\sigma}$ to generate the matrix $\bf{H_{\sigma}}$ and then taking the product of each the non-zero elements of $\bf{H_{\sigma}}$ 
with total non-zero elements of identity matrix ${\cal{I}}_{\overline{\sigma}}$ which is equal to $count_{\overline{\sigma}}$; where if $\sigma=\uparrow(\downarrow)$ then $\overline{\sigma}=\downarrow(\uparrow)$.

\section{Performance}

Table II(Table III) shows the comparision of time taken to generate the
non-diagonal part of the Hamiltonian matrix by applying the Hamiltonian to
two-spin basis states $I^{,}s$ and one-spin basis states $I_{\sigma}^{,}s$ respectively 
in 1D system (4 $\times$ 4 lattice).
We find that by using one spin configuration basis the total CPU time for computation of non-diagonal part of Hamiltonian matrix
is reduced to 1/9 $\sim$ 1/11 compared to the two spin configuration basis.
While computing the non-diagonal part of the Hamiltonian matrix, the action of a hopping term changes a basis state to another whose index
can be found by performing a binary search without storing the respective indices of the basis states in a seperate array.
Table IV shows the comparision of time taken to compute the ground state eigenvalue and eigenvector by using binary search to find the 
index of a given basis state to generate the Hamiltonian matrix to implement the
Lanczos algorithm by working with two-spin basis states $I^{,}s$ and one-spin basis states $I_{\sigma}^{,}s$
respectively. We find that by using one spin configuration basis, the total CPU time for computation of ground state eigenvalue and eigenvector
is reduced to 1/3 $\sim$ 1/5 compared to
the two spin configuration basis. From column 5 and column 6 of table IV, we find that using one-spin configuration basis states, 
the time taken is almost the same while performing a binary
search to find the index of a given basis state and seperately storing the indices of the basis states respectively.
Thus we find that with one-spin configuration basis states,
the binary search becomes as effective as separately storing the indices of the basis states.
The simplest version of Lanczos algorithm without any form of re-orthogonalization has been implemented
to compute the eigenvalues and eigenvectors given in table IV.

\begin{table}
\begin{tabular}{|c|c|c|c|c|c|}
	\hline
$M$  & $N_{\uparrow}$ & $N_{\downarrow}$   & Hilbert Space & Time       &   Time           \\
     &                &                    &  Dimension    & Two-spin   &  One-spin         \\
	\hline
16   &  8             &      8             & 165 636 900   & 78.66 s    &  8.74 s           \\
14   &  7             &      7             &  11 778 624   &  5.02 s    &  0.55 s            \\
        \hline
\end{tabular}
\caption{Time taken to generate the non-diagonal part of the 1-D Hubbard model
with $U=4$ and $t=1$(half-filled, $S^{z}=0$)on an intel i7 processor 
machine by applying the Hamiltonian to the two-spin basis states $I^{,}s$ and 
one-spin basis states $I_{\sigma}^{,}s$ respectively.}
\end {table}

\begin{table}
\begin{tabular}{|c|c|c|c|c|c|c|}
      \hline
$M$  & $N_{\uparrow}$ &  $N_{\downarrow}$   & Hilbert Space      &   Time      &    Time\\
     &                &                     & Dimension          & two-spin    &  one-spin\\
       \hline
16   &    8           &      8              &  165 636 900       & 186.71 s   &    18.32 s\\
16   &    7           &      7              &  130 873 600       & 146.02 s   &    14.24 s\\
      \hline
\end{tabular}
\caption{Time taken to generate the non-diagonal part of the 4$\times$4 square lattice
with $U=4$ and $t=1$ on an intel i7 processor 
machine by applying the Hamiltonian to the two-spin basis states $I^{,}s$ and 
one-spin basis states $I_{\sigma}^{,}s$ respectively.}
\end{table}

\begin{table}[!]
\begin{tabular}{|c|c|c|c|c|c|}
      \hline
$M$  &  Hilbert space       &     $E_{min}$     &           Time       &    Time          &  Time          \\
     &  Dimension           &                   &         two-spin     &  one-spin        &  one-spin         \\
     &                      &                   &          [Binary     &  [Binary         &  [Storing        \\
     &                      &                   &          Search]     &  Search]         &  Indices]         \\
       \hline
16   &  165 636 900         &   -9.2144309716   &         60427.63 s   &    16245.27 s    &  16190.21 s          \\
14   &   11 778 624         &   -8.0883491038   &          3459.85 s   &      702.15 s    &    701.42 s          \\
      \hline
\end{tabular}
\caption{Time taken to compute ground state eigenvalue and eigenvector of a 1-D Hubbard model
with $U=4$ and $t=1$(Half-filled, $S^{z}=0$)on an intel i7 processor
machine using Lanczos algorithm that includes on-the-fly generation of the Hamiltonian matrix 
using binary search to find the index of a given basis state while
working on the two-spin basis states $I^{,}s$ in column 4 and one-spin basis states $I_{\sigma}^{,}s$ in column 5 respectively
and storing the indices of basis states separately while working on one-spin basis states $I_{\sigma}^{,}s$ in column 6 .}
\end{table}

\section{Using translation symmetry}

The Hamiltonian (\ref{count1}) commutes with the translation operator $\colon$ $Tc^{\dagger}_{i,\sigma}\rightarrow c^{\dagger}_{i+1,\sigma}$ incorporated by imposing the periodic boundary condition; (i.e. N+1 $\equiv$ 1, N+2 $\equiv$ 2 for a linear chain). Using the translation symmetry, the size of {\bf{H}} can be further reduced by a factor of $M$ equal to the number of sites\cite{aw2010,aw2013}.

The set of all translationally related states form a cycle and the number of distinct elements of a cycle is called its period P. The period of a cycle is either equal to the total number of sites $M$ or any factor of $M$.

The eigenvalue of the translation operator $T$ is $e^{ik}$ with corresponding eigenvector $\left|a(k)\right\rangle=1/\sqrt{P} \sum_{r=0}^{P-1}e^{-ikr}T^{r}\left|a\right\rangle$, where $k=2\pi s/P$, $s=-P/2+1,...,P/2$ and $\left|a\right\rangle$ is a reference state of the 
cycle of period $P$. Usually a state having the smallest integer value among all the members of the cycle is chosen as a representative state.

For simplicity, we implement our scheme for $k=0$ space.

For a given spin $\sigma (\uparrow or \downarrow)$ with given $N_{\sigma}$ and $M$, a number $z$ is a period only if:
\begin{equation}
\textbf{mod}(M,z)=0\\ \label{count4}
\end{equation}
\begin{equation}
\textbf{mod}(N_{\sigma},M/z)=0 \label{count5}
\end{equation}
with total number of representative states equal to
\begin{equation} 
\frac{{^{z}C}_{y_{z}}-\sum_{i}{^{i}C}_{y_{i}}}{z}\label{count6},
\end{equation}
where $y_{b}=\frac{N_{\sigma}b}{M}$ and $i$($\textless z$) is a period of spin configuration $\sigma (\uparrow or \downarrow)$ with given $N_{\sigma}$ and $M$ satisfying $\textbf{mod}(z,i)=0$.

The combined representative state $\left|r\right\rangle$ representing both spin $\uparrow$ and $\downarrow$ configurations is given by
\begin{equation} 
\left|r\right\rangle=\left|r_{\uparrow}\right\rangle T^{n} \left|r_{\downarrow}\right\rangle \label{count7},
\end{equation} 
where $\left|r_{\uparrow}\right\rangle(\left|r_{\downarrow}\right\rangle)$
is spin up(down) representative state and $n$
varies from $0$ to $f-1$, wherein $f$ is the highest common factor (h.c.f.) of periods of $\left|r_{\uparrow}\right\rangle$ and $\left|r_{\downarrow}\right\rangle$. Table I shows the formation of the combined representative states from $\left|r_{\uparrow}\right\rangle$ and $\left|r_{\downarrow}\right\rangle$ respectively.

\begin{table}
\begin{tabular}{|c|c|c|c|c|}
	\hline
$\left|r_{\uparrow}\right\rangle$  & $\left|r_{\downarrow}\right\rangle$ & $n$ & $T^{n}\left|r_{\downarrow}\right\rangle$  & $Index$ \\
conf.& conf.&  &  &  \\
	\hline
0011 & 0001 & 0 & 0001 & 1 \\
0011 & 0001 & 1 & 1000 & 2 \\
0011 & 0001 & 2 & 0100 & 3 \\
0011 & 0001 & 3 & 0010 & 4 \\
0101 & 0001 & 0 & 0001 & 5 \\
0101 & 0001 & 1 & 1000 & 6 \\
\hline
\end{tabular}
\caption{Spin up configurations of the representative states(col.I), spin down configurations of the representative states(col.II), translation of spin down states(col.III), translated spin down states(col.IV) and index of combined representative states(col.V) for $M=4$ in a $sector(N_{\uparrow}=2,N_{\downarrow}=1)$.}
\end{table}

Each combined representative state expressed in Eq.(\ref{count7}) and all its translations follows a relation:
\begin{equation}
translation_{\uparrow}-translation_{\downarrow}+n=x_{\downarrow}period_{\downarrow}-x_{\uparrow}period_{\uparrow}, \label{count8}
\end{equation}
where $translation_{\uparrow}$($translation_{\downarrow}$) is the translation of spin up(down) basis state with respect to its representative state, where $x_{\uparrow}$($x_{\downarrow}$) is an integer varying from 0 to f(g), wherein $f=l.c.m.(period_{\uparrow},period_{\downarrow})/period_{\uparrow}-1$ and
$g=l.c.m.(period_{\uparrow},period_{\downarrow})/period_{\downarrow}$; $l.c.m.$ denotes the lowest common multiple.

In our scheme, we do not generate the combined representative states or any of their translations. We \textbf{do not} work with the combined representative states given by Eq.(\ref{count7}).

We find all the possible periods of a spin $\sigma (\uparrow or \downarrow)$ configuration by checking all numbers equal to or less than $M$ satisfying Eq.(\ref{count4}) and Eq.(\ref{count5}) and store them in an integer array $period_{\sigma}(1:perep_{\sigma})$, where $perep_{\sigma}$ is the total number of periods. We also compute the total number of representative states for a given period using Eq.(\ref{count6}) and store them in an integer array $nstate_{\sigma}(1:perep_{\sigma}$). The dimension of the Hilbert space i.e., the total number of combined representative states expressed in Eq.(\ref{count7}) is
\begin{equation}
\sum_{i=1}^{perep_{\uparrow}}alls(i)nstate_{\uparrow}(i),
\end{equation} 
where $alls(i)=\sum_{j=1}^{perep_{\downarrow}}nstate_{\downarrow}(j)[h.c.f.(period_{\uparrow}(i),$\\
$period_{\downarrow}(j))]$.
We store each of the representative states of a spin $\sigma (\uparrow or \downarrow)$ configuration periodwise in a two-dimensional array $repres_{\sigma}(i,j)$, where $i$ is the index of its period and $j$ is the index of the representative state within the period.

To compute the non-diagonal part of the Hamiltonian matrix of a spin $\sigma (\uparrow or \downarrow)$ configuration,
we generate all the $count_{\sigma}$ $basis_{\sigma}$ and store the index of their period, index of representative state in that period and translation with respect to the representative state in one-dimensional arrays $per_{\sigma}$, $rep_{\sigma}$ and $trans_{\sigma}$ respectively; these arrays take the integral value of $basis_{\sigma}$ and return the corresponding quantities. We also store the sign (phase factor) of each $basis_{\sigma}$(defined by index of  its period, index of representative state within the period, and translation with respect to representative state) obtained after translations with respect to the representative state to get that particular state in a one-dimensional integer array $esign_{trans_{\sigma}}$ (shown in Table II). For each representative state(defined by the index of its period and index of the representative state within the period) the phase factor obtained after period number of translations to get back the same state, $T^{P}repres_{\sigma}(i,j)\rightarrow repres_{\sigma}(i,j)$ is also stored in a one-dimensional array $esignb_{\sigma}$. For example, $T^{2}\left|0101\right\rangle=-\left|0101\right\rangle$. The array $esignb_{\sigma}$ is stored for both spin $\uparrow$ and $\downarrow$ configurations.\\
Let the action of a spin $\sigma (\uparrow or \downarrow)$ hopping term change a representative state $repres_{\sigma}(i,j)$ to $T^{s}repres_{\sigma}(p,l)$,
\begin{equation}
-t(c_{i\sigma}^{\dagger}c_{j\sigma}+c_{j\sigma}^{\dagger}c_{i\sigma})os=-t(esign_{oper})ns,
\end{equation}
where $repres_{\sigma}(i,j)\equiv os$, $T^{s}repres_{\sigma}(p,l)\equiv ns$ and $esign_{oper}$ takes care of the sign as explained earlier. All the elements of {\bf{H}} due to this particular transition(spin $\sigma$ hopping) and all translations of this transition are generated using Algorithm III given in Appendix \ref{algo3} for $\sigma=\uparrow$ and Algorithm IV given in Appendix \ref{algo4} for $\sigma=\downarrow$, where  $matrix(r,s)$ is the (r,s)th element of {\bf{H}}, the matrix elements corresponding to the two values of $esign=1$ and $esign=-1$ are defined as $matel(1)=-t$ and $matel(-1)=t$ respectively($t$ is the hopping amplitude). 
The above procedure is repeated for each of the spin $\sigma$ hopping terms acting on each of the spin $\sigma$ 
representative states to obtain all the matrix elements due to spin $\sigma$ hopping.

To compute the diagonal part of {\bf{H}}, we use the representative states of spin up configuration and the representative states of spin down configuration as shown in Algorithm V given in Appendix \ref{algo5}, where $matrix(r,s)$ is the (r,s)th element of {\bf{H}} and $U$ is the onsite repulsion.
 
\begin{table}
\begin{tabular}{|c|c|c|c|c|}
	\hline
$basis_{\sigma}$  & $per_{\sigma}$ & $rep_{\sigma}$ & $trans_{\sigma}$  & $esign_{trans_{\sigma}}$ \\
 conf.&  &  &  &  \\
	\hline
0011 & 1 & 1 & 0 & 1 \\
1001 & 1 & 1 & 1 & -1 \\
1100 & 1 & 1 & 2 & 1 \\
0110 & 1 & 1 & 3 & 1   \\
0101 & 2 & 1 & 0 & 1 \\
1010 & 2 & 1 & 1 & -1 \\
\hline
\end{tabular}
\caption{Spin configurations of $basis_{\sigma}$(col.I), index of their period(col.II), index of their representative state within the period(col.III), translation with respect to representative state(col.IV) and phase factor with respect to representative state(col.V) for $M=4$ and $N_{\sigma}=2$.}
\end{table}

The one-particle Green's function carries information about single particle excitation in the (interacting) system. The spectral function obtained from the Green's function gives the distribution of single-particle states required in calculation of any of the
transport properties, equilibrium as well as non-equilibrium. In our scheme, the Green's function can be computed using the basis states of one spin $\sigma (\uparrow or \downarrow)$ configuration at a time without generating the complete basis states representing both spin configurations.

\subsection{Green's function at zero temperature}
he calculation of dynamical 
properties of a given Hamiltonian is done using Lanczos technique by constructing
a full continued fraction\cite{hh1975,ed1994,pj2011}.
The zero temperature Green's function is expressed as:
\begin{equation}
G_{\sigma}(p,i\omega_{n})=
\left\langle\psi_{0}\right|c_{p,\sigma}\frac{1}{i\omega_{n}+(E_{0}-{\bf{H}})}c^{\dagger}_{p,\sigma}\left|\psi_{0}\right\rangle+\nonumber
\end{equation}
\begin{equation}
\left\langle\psi_{0}\right|c^{\dagger}_{p,\sigma}\frac{1}{i\omega_{n}-(E_{0}-{\bf{H}})}c_{p,\sigma}\left|\psi_{0}\right\rangle \label{count8}
\end{equation}
where ${\bf{H}}$ is the Hamiltonian matrix, $E_{0}$ and $\psi_{0}$ are the groundstate eigenvalue and eigenvector of ${\bf{H}}$ respectively.
Here we are required to use
$ c^{\dagger}_{p,\sigma}\left|\psi_{0}\right\rangle$ 
as the starting vector in the Lanczos procedure. The vector $c^{\dagger}_{p,\uparrow}\left|\psi_{0}\right\rangle \equiv vecin $ is computed from the groundstate eigenvector of the Hamiltonian matrix of sector ($N_{\uparrow},N_{\downarrow}$) for spin up Green's function using the following algorithm:\\\\

\noindent$vecin(1:count1_{\uparrow}count_{\downarrow})=0$\\
\noindent\textbf{for} $n=1:count_{\uparrow}$\\
\indent\indent\textbf{if} site $p$ of $basis_{\uparrow}(n)$ is unoccupied\\ 
\indent\indent\indent\indent$\lbrace c^{\dagger}_{\uparrow,p}basis_{\uparrow}(n)\rightarrow (esign)newstate \rbrace $\\
\indent\indent\indent\indent$r=index1_{\uparrow}(newstate)$\\
\indent\indent\indent\indent\textbf{for} $s=1:count_{\downarrow}$\\
\indent\indent\indent\indent\indent\indent$f=(s-1)count_{\uparrow}+n$\\
\indent\indent\indent\indent\indent\indent$g=(s-1)count1_{\uparrow}+r$\\    
\indent\indent\indent\indent\indent\indent$vecin(g)=vecin(g)+(esign)gndvec(f)$\\
\indent\indent\indent\indent\textbf{end}\\   
\indent\indent\textbf{end}\\                              
\textbf{end},\\

where $count_{\uparrow}$($count1_{\uparrow}$) is the total number of spin up basis states for $N_{\uparrow}$($N_{\uparrow}+1$) electrons, $esign$ takes care of the sign as explained earlier and the array $index1_{\uparrow}$ stores the indices of ($N_{\uparrow}+1$) spin up electrons basis states. $gndvec(1:count_{\uparrow}count_{\downarrow})$ is the groundstate eigenvector of the Hamiltonian matrix of sector ($N_{\uparrow},N_{\downarrow}$).

\subsection{Green's function at finite temperature}
At finite temperature, the Green's function is computed using the expression:
 
\begin{equation}
G_{\sigma}(p,i\omega_{n})=\sum_{i,j}
\frac{\left|\left\langle i|c^{\dagger}_{p,\sigma}|j\right\rangle\right|^{2}}{E_{j}-E_{i}+i\omega_{n}}\frac{(e^{-\beta E_{i}}+e^{-\beta E_{j}})}{Z}\label{count9},
\end{equation}

where $c_{p \sigma}^{\dagger}$ is fermion creation operator at site $p$ with spin $\sigma (\uparrow or \downarrow)$, $Z$ is the partition function and
$\omega_{n}$ is the matsubara frequency. The full set of states $\left|i\right\rangle(\left|j\right\rangle)$ are the eigenvectors with corresponding eigenvalues $E_{i}(E_{j})$ belonging to the sector having $(N_{\sigma}$ and $N_{\overline{\sigma}}$($N_{\sigma}+1$ and $N_{\overline{\sigma}}$) electrons, where if $\sigma=\uparrow(\downarrow)$ then $\overline{\sigma}=\downarrow(\uparrow)$ and $N_{\sigma}(N_{\overline{\sigma}})$ varies from $0$($0$) to $M-1$($M$). We compute the quantity $\left|\left\langle i|c^{\dagger}_{p,\uparrow}|j\right\rangle\right|^{2}\equiv mat2_{\uparrow}$ involving two sectors $(N_{\uparrow},N_{\downarrow})$ and $(N_{\uparrow}+1,N_{\downarrow})$ for spin up Green's function via the following algorithm:\\\\

\noindent$mat_{\uparrow}=0$\\
\textbf{for} $n=1,count_{\uparrow}$\\
\indent\indent\textbf{if} site $p$ of $basis_{\uparrow}(n)$ is unoccupied\\
\indent\indent\indent\indent$\lbrace c^{\dagger}_{\uparrow,p}basis_{\uparrow}(i)\rightarrow (esign)newstate \rbrace $\\
\indent\indent\indent\indent$r=index1_{\uparrow}(newstate)$\\
\indent\indent\indent\indent\textbf{for} $s=1:count_{\downarrow}$\\
\indent\indent\indent\indent\indent\indent$f=(s-1)count_{\uparrow}+n$\\
\indent\indent\indent\indent\indent\indent$g=(s-1)count1_{\uparrow}+r$\\ 
\indent\indent\indent\indent\indent\indent$mat_{\uparrow}=mat_{\uparrow}+(esign)zr(f,j)zrr(g,i)$\\   
\indent\indent\indent\indent\textbf{end}\\
\indent\indent\textbf{end}\\
\textbf{end}\\
$mat2_{\uparrow}=mat_{\uparrow}mat_{\uparrow}$,\\
     
where $count_{\uparrow}$($count1_{\uparrow}$) is the total number of spin up basis states for $N_{\uparrow}$($N_{\uparrow}+1$) electrons, $esign$ takes care of the sign as explained earlier and the array $index1_{\uparrow}$ stores the indices of $(N_{\uparrow}+1)$ spin up electrons basis states. $zr(f,j)$($zrr(g,i)$) is the $f^{th}$($g^{th}$) component of the eigenvector corresponding to the $j^{th}$($i^{th}$) eigenvalue of the Hamiltonian matrix of sector $(N_{\uparrow},N_{\downarrow})$($(N_{\uparrow}+1,N_{\downarrow})$).

\section{Applications to the Dynamical Mean Field Theory}
In practice, the most difficult step in the DMFT iterative procedure is the repeated calculation of the impurity Green's function:
\begin{equation}
G_{imp}(i\omega_{n})=\int_{0}^{\beta}d\tau
e^{i\omega_{n}\tau}(-\left\langle T_{
\tau}c(\tau)c(0)^{\dagger}\right\rangle)
\end{equation} 
required in the DMFT self consistency loop\cite{ag1992}.
The advantage of our scheme in DMFT is twofold:
Firstly, DMFT maps the Hubbard model onto Anderson impurity model.
Algorithm II is much simplified for the Anderson impurity model(AIM)\cite{pw1961}, (where the onsite Coulomb interaction is only on one impurity 
site) which is solved using ED to generate the eigenvalues and eigenvectors.
Secondly, using these eigenvalues and eigenvectors, the Green's function is computed using Eq.(\ref{count8}) or Eq.(\ref{count9}) efficiently with our
scheme as discussed in the previous section. Even Green's function at very low temperature computed by the set of equations used by Capone et al\cite{mc2007} is also targeted profitably with our scheme\cite{ms2014}.

\section{Extention to other models}
The present scheme can be applied to the Anderson lattice model with an extra simplification that the onsite Coulomb interaction is only among the correlated electrons
and not among the uncorrelated conduction electrons. Pariser-Parr-Popple model\cite{rr1953,ja1953} that contains additional intersite Coulomb repulsion and also the one-band extended Hubbard model can be dealt with using both $I_{\uparrow}$ and $I_{\downarrow}$ {\bf{without generating the $I^{,}s$}}. This is much easier for some specific sectors, for example, for $N_{\uparrow}=N_{\downarrow}$ the basis states for both spin $\sigma (\uparrow or \downarrow)$ configurations are the same.
For half filled $(N_{\uparrow}+N_{\downarrow}=M)$ sectors, the basis 
states spanning ${\cal{V}}^{N_{\sigma}}$ can be obtained from the basis states spanning ${\cal{V}}^{N_{\overline{\sigma}}}$, where if $\sigma=\uparrow(\downarrow)$ then $\overline{\sigma}=\downarrow(\uparrow)$ using the following loop:

\noindent$g=0$\\
\noindent\textbf{for} $i=count_{\sigma}:-1:f$\\
\indent\indent $ g=g+1 $\\
\indent\indent $ oo=basis_{\sigma}(i) $\\
\indent\indent $ basis_{\sigma}(i)=\textbf{not}(basis_{\sigma}(g))$\\
\indent\indent $ basis_{\sigma}(g)=\textbf{not}(oo)$\\
\textbf{end}

where the function $\textbf{not}(k)$ returns the logical compliments of the bits of integer $k$ and $f=\frac{count_{\sigma}}{2}+1$ if $count_{\sigma}$ is even and $f=\frac{count_{\sigma}+1}{2}$ if $count_{\sigma}$ is odd.\\

\section{Conclusions}
The scheme of working with $I_{\sigma}^{,}s$ spanning ${\cal{V}}^{N_{\sigma}}$ at a time 
$\sigma(\uparrow or\downarrow)$ to generate the full Hamiltonian matrix of the one-band Hubbard model that we 
present above has the advantages listed below as compared to working with the complete basis states $I^{,}s$ spanning the full Hilbert space ${\cal{V}}^{(N_{\uparrow},N_{\downarrow})}={\cal{V}}^{N_{\uparrow}}\otimes{\cal{V}}^{N_{\downarrow}}$:

$\bullet$We compute and store $I_{\sigma}^{,}s$ for a given spin $\sigma(\uparrow or\downarrow)$ at a time instead of computing and 
storing $I^{,}s$, reducing both storage
and CPU time requirements. For the specific sector $N_{\uparrow}=N_{\downarrow}$, i.e. $S^{z}=0$, the 
basis states of only one spin configuration $\sigma(\uparrow or\downarrow)$ are required to be computed.

$\bullet$To generate the non-diagonal part of the Hamiltonian matrix, we apply 
the spin up hopping terms to $count_{\uparrow}$ spin up basis states spanning ${\cal{V}}^{N_{\uparrow}}$ and the 
spin down hopping terms to $count_{\downarrow}$ spin down basis states spanning ${\cal{V}}^{N_{\downarrow}}$ instead of applying the hopping terms of both spin configurations to $count_{\uparrow}count_{\downarrow}$ basis states spanning ${\cal{V}}^{(N_{\uparrow},N_{\downarrow})}$, saving significant computation time. The total CPU time required to generate the non-diagonal matrix elements of the Hamiltonian is found to reduce by a factor of 10 for our test runs.

$\bullet$ The storage requirement of the non-diagonal part of the sparse full Hamiltonian matrix of
effective dimension $(count_{\uparrow}count_{\downarrow})\times F$
is reduced to the storage requirement of two small hermitian matrices of
effective dimensions $count_{\uparrow}\times(F/2)$ due to spin up hopping and
$count_{\downarrow}\times(F/2)$ due to spin down hopping respectively, where
$F(<<count_{\sigma})$ is the maximum number of non-zero off-diagonal elements in any given row of the Hamiltonian matrix.
For the $S^{z}=0$ sector, the two small hermitian matrices 
are \textbf{identical}, thereby requiring storage only for one matrix.

$\bullet$ We need to store or search the indices of the basis states of one spin
configuration at a time and not to bother about the indices of the basis states
of the other spin configuration\cite{ds2010}.A binary search requires at the  
most $O(log_{2}(count_{\uparrow}))$ or $O(log_{2}(count_{\downarrow}))$ 
comparisions to find the index of a given $I_{\uparrow}$ or $I_{\downarrow}$ respectively
as compared to $O(log_{2}(count_{\uparrow}count_{\downarrow}))$ comparisions to find the index of a given $I$.
This amounts to a huge gain in computation time while generating the Hamiltonian matrix on the fly required for matrix vector multiplication
for diagonalization.
For $S^{z}=0$ or half filled ($N_{\uparrow}+N_{\downarrow}=M$)
sectors, where $count_{\uparrow}=count_{\downarrow}$, the maximum number of comparisions required in 
binary search for finding a particular index would be of order half while working with $I_{\uparrow}^{,}s$ or $I_{\downarrow}^{,}s$ than those
with $I^{,}s$.

$\bullet$ Working with $I^{,}s$ and then extracting $I_{\uparrow}$ and
$I_{\downarrow}$ from $I$ by examining its bits\cite{aj2008} is done away with by
working with only $I_{\uparrow}^{,}s$ and $I_{\downarrow}^{,}s$ at a time.

$\bullet$Our scheme is inherently parallelizable and can be profitably implemented on a parallel machine
reducing inter-node communication significantly.\\\\

The present scheme being readily parallelizable and economical in terms of CPU time and memory has varied applications:
\begin{itemize}
\item Can be implemented on translationally symmetric lattices.
\item Can be employed to find the static and dynamical properties like the correlation functions of the model systems.
\item Makes ED a powerful impurity solver in DMFT scheme.
\item Can be extended to other lattice models in condensed matter, molecular and quantum chemistry systems.
\end{itemize}

\begin{acknowledgements}
Medha Sharma is thankful to DST for financial assistance in the form
of Inspire Fellowship. Medha Sharma also thanks S.R. Hassan and Rajesh
Karan for introducing her to the field of DMFT and Centre for Theoretical Physics,
Jamia Millia Islamia (New Delhi, India) for providing computational facility.
\end{acknowledgements}

\appendix
\section{Algorithm I}
\label{algo1}
\noindent $minrange=0$\\
$maxrange=0$\\
\textbf{for} $i=1:N_{\sigma}$\\
\indent\indent$minrange=minrange+2^{i-1}$\\
\indent\indent$maxrange=maxrange+2^{M-i}$\\
\textbf{end}\\
$count_{\sigma}=0$\\
\textbf{for} $i=minrange:maxrange$\\
\indent\indent$nbit=0$\\
\indent\indent \textbf{for} $j=0:(M-1)$\\
\indent\indent\indent\indent \textbf{if} \textbf{bittest}$(i,j)$\\
\indent\indent\indent\indent\indent\indent $nbit=nbit+1$\\
\indent\indent\indent\indent \textbf{end}\\
\indent\indent \textbf{end}\\
\indent\indent\textbf{if} $nbit=N_{\sigma}$\\
\indent\indent\indent\indent $count_{\sigma}=count_{\sigma}+1$\\
\indent\indent\indent\indent $basis_{\sigma}(count_{\sigma})=i$\\
\indent\indent\indent\indent $index_{\sigma}(i)=count_{\sigma}$\\
\indent\indent\textbf{end}\\
\textbf{end}

\section{Algorithm II}
\label{algo2}
\noindent $matrix(1:count_{\uparrow}count_{\downarrow},1:count_{\uparrow}count_{\downarrow})=0$\\
$\lbrace$Using Algorithm I compute $basis_{\uparrow}(1:count_{\uparrow})$ and store in $basis_{\sigma}(1:count_{\uparrow})\rbrace$\\
\textbf{for} $i=1:M$\\
\indent\indent$point_{\uparrow}=0$\\
\indent\indent\textbf{for} $j=1:count_{\uparrow}$\\
\indent\indent\indent\indent\textbf{if} $N_{\uparrow}+N_{\downarrow}\leqslant M$\\
\indent\indent\indent\indent\indent\indent\textbf{if} site $i$ of $basis_{\sigma}(j)$ is occupied\\
\indent\indent\indent\indent\indent\indent\indent\indent $point_{\uparrow}=point_{\uparrow}+1$\\
\indent\indent\indent\indent\indent\indent\indent\indent $index_{U\uparrow}(i,point_{\uparrow})=j$\\
\indent\indent\indent\indent\indent\indent\indent\indent\textbf{if} $N_{\uparrow}=N_{\downarrow}$\\
\indent\indent\indent\indent\indent\indent\indent\indent\indent\indent $point_{\downarrow}=point_{\uparrow}$\\
\indent\indent\indent\indent\indent\indent\indent\indent\indent\indent $index_{U\downarrow}(i,point_{\downarrow})=j$\\
\indent\indent\indent\indent\indent\indent\indent\indent\textbf{end}\\
\indent\indent\indent\indent\indent\indent\textbf{end}\\
\indent\indent\indent\indent\textbf{else}\\
\indent\indent\indent\indent\indent\indent\textbf{if} site $i$ of $basis_{\sigma}(j)$ is unoccupied\\
\indent\indent\indent\indent\indent\indent\indent\indent $point_{\uparrow}=point_{\uparrow}+1$\\
\indent\indent\indent\indent\indent\indent\indent\indent $index_{U\uparrow}(i,point_{\uparrow})=j$\\
\indent\indent\indent\indent\indent\indent\indent\indent\textbf{if} $N_{\uparrow}=N_{\downarrow}$\\
\indent\indent\indent\indent\indent\indent\indent\indent\indent\indent $point_{\downarrow}=point_{\uparrow}$\\
\indent\indent\indent\indent\indent\indent\indent\indent\indent\indent $index_{U\downarrow}(i,point_{\downarrow})=j$\\
\indent\indent\indent\indent\indent\indent\indent\indent\textbf{end}\\
\indent\indent\indent\indent\indent\indent\textbf{end}\\
\indent\indent\indent\indent\textbf{end}\\
\indent\indent\textbf{end}\\
\textbf{end}\\
\textbf{if} $N_{\uparrow} \neq N_{\downarrow}$\\
\indent\indent $\lbrace$Using Algorithm I compute $basis_{\downarrow}(1:count_{\downarrow})$ and store in $basis_{\sigma}(1:count_{\downarrow})\rbrace$\\
\indent\indent \textbf{for} $i=1:M$\\
\indent\indent\indent\indent$point_{\downarrow}=0$\\
\indent\indent\indent\indent\textbf{for} $j=1:count_{\downarrow}$\\
\indent\indent\indent\indent\indent\indent\textbf{if} $N_{\uparrow}+N_{\downarrow}\leqslant M$\\
\indent\indent\indent\indent\indent\indent\indent\indent\textbf{if} site $i$ of $basis_{\sigma}(j)$ is occupied\\
\indent\indent\indent\indent\indent\indent\indent\indent\indent\indent $point_{\downarrow}=point_{\downarrow}+1$\\
\indent\indent\indent\indent\indent\indent\indent\indent\indent\indent $index_{U\downarrow}(i,point_{\downarrow})=j$\\
\indent\indent\indent\indent\indent\indent\indent\indent\textbf{end}\\
\indent\indent\indent\indent\indent\indent\textbf{else}\\
\indent\indent\indent\indent\indent\indent\indent\indent\textbf{if} site $i$ of $basis_{\sigma}(j)$ is unoccupied\\
\indent\indent\indent\indent\indent\indent\indent\indent\indent\indent $point_{\downarrow}=point_{\downarrow}+1$\\
\indent\indent\indent\indent\indent\indent\indent\indent\indent\indent $index_{U\downarrow}(i,point_{\downarrow})=j$\\
\indent\indent\indent\indent\indent\indent\indent\indent\textbf{end}\\
\indent\indent\indent\indent\indent\indent\textbf{end}\\
\indent\indent\indent\indent\textbf{end}\\
\indent\indent\textbf{end}\\
\textbf{end}\\
\textbf{if} $N_{\uparrow}+N_{\downarrow}>M$\\
\indent\indent $g=U(N_{\uparrow}+N_{\downarrow}-M)$\\
\indent\indent\textbf{for} $i=1:count_{\uparrow}count_{\downarrow}$\\
\indent\indent\indent\indent $matrix(i,i)=matrix(i,i)+g$\\
\indent\indent\textbf{end}\\
\textbf{end}\\
\textbf{if} $N_{\uparrow}=N_{\downarrow}$\\
\indent\indent\textbf{if} $N_{\uparrow}+N_{\downarrow}\leqslant M$\\
\indent\indent\indent\indent $g=UN_{\uparrow}$\\
\indent\indent\textbf{else}\\
\indent\indent\indent\indent $g=U(M-N_{\uparrow})$\\
\indent\indent\indent\indent\textbf{for} $i=1:count_{\uparrow}$\\
\indent\indent\indent\indent\indent\indent $k=(count_{\uparrow}+1)i-count_{\uparrow}$\\
\indent\indent\indent\indent\indent\indent $matrix(k,k)=matrix(k,k)+g$\\
\indent\indent\indent\indent\textbf{end}\\
\indent\indent\textbf{end}\\
\textbf{end}\\
\textbf{for} $i=1:M$\\
\indent\indent\textbf{for} $k=1:point_{\uparrow}$\\
\indent\indent\indent\indent\textbf{for} $l=1:point_{\downarrow}$\\
\indent\indent\indent\indent\indent\indent\textbf{if} $N_{\uparrow}=N_{\downarrow}$\\
\indent\indent\indent\indent\indent\indent\indent\indent\textbf{if} $l\neq k $\\
\indent\indent\indent\indent$r=(index_{U\downarrow}(i,l)-1)count_{\uparrow}+index_{U\uparrow}(i,k)$\\
\indent\indent\indent\indent$matrix(r,r)=matrix(r,r)+ U$\\
\indent\indent\indent\indent\indent\indent\indent\indent\textbf{end}\\
\indent\indent\indent\indent\indent\indent\textbf{else}\\
\indent\indent\indent\indent$r=(index_{U\downarrow}(i,l)-1)count_{\uparrow}+index_{U\uparrow}(i,k)$\\
\indent\indent\indent\indent$matrix(r,r)=matrix(r,r)+ U$\\
\indent\indent\indent\indent\indent\indent\textbf{end}\\ 
\indent\indent\indent\indent\textbf{end}\\
\indent\indent\textbf{end}\\
\textbf{end}

\section{Pseudocode}
\label{pseudocode1}
\noindent\textbf{!MASTER PART}\\
$\lbrace$ Store the matrix ${\bf{H_{\uparrow}}}$ of dimension $count_{\uparrow}$ on each node in sparse matrix format.$\rbrace $\\
\textbf{!}Split the vector $q_{old}$ into $count_{\downarrow}$ blocks of dimension \\
\textbf{!}$count_{\uparrow}$ each and distribute among all the nodes\\
\textbf{!}($n_{p}$ is the total number of nodes).\\
\textbf{if} $\textbf{mod}(count_{\downarrow},n_{p})=0$\\
\textbf{!}Store $count_{\downarrow}/n_{p}$ blocks of vector $q_{old}$ on each node.\\
\indent\indent$r=count_{\downarrow}/n_{p}$\\
\indent\indent$a=0$\\
\indent\indent\textbf{for} $i_{p}=1:n_{p}$\\
\indent\indent\indent\indent\textbf{for} $j=1:r$\\
\indent\indent\indent\indent\indent\indent\textbf{for} $i=1:count_{\uparrow}$\\
\indent\indent\indent\indent\indent\indent\indent\indent$a=a+1$\\
\indent\indent\indent\indent\indent\indent\indent\indent$Q(i,j)=q_{old}(a)$\\
\indent\indent\indent\indent\indent\indent\indent\indent$\lbrace$ Store $Q(i,j)$ on $i_{p}$ node.$\rbrace$\\
\indent\indent\indent\indent\indent\indent\textbf{end}\\
\indent\indent\indent\indent\textbf{end}\\
\indent\indent\textbf{end}\\
\textbf{else}\\
\textbf{!}Store $(count_{\downarrow}/n_{p}+1)$ blocks of vector $q_{old}$ on \\
\textbf{!}$\textbf{mod}(count_{\downarrow},n_{p})$ nodes.\\
\indent\indent$z=\textbf{mod}(count_{\downarrow},n_{p})$\\
\indent\indent$r=count_{\downarrow}/n_{p}$\\
\indent\indent$s=r+1$\\
\indent\indent$a=0$\\
\indent\indent\textbf{for} $i_{p}=1:z$\\
\indent\indent\indent\indent\textbf{for} $j=1:s$\\
\indent\indent\indent\indent\indent\indent\textbf{for} $i=1:count_{\uparrow}$\\
\indent\indent\indent\indent\indent\indent\indent\indent$a=a+1$\\
\indent\indent\indent\indent\indent\indent\indent\indent$Q(i,j)=q_{old}(a)$\\
\indent\indent\indent\indent\indent\indent\indent\indent$\lbrace$ Store $Q(i,j)$ on $i_{p}$ node.$\rbrace$\\
\indent\indent\indent\indent\indent\indent\textbf{end}\\
\indent\indent\indent\indent\textbf{end}\\
\indent\indent\textbf{end}\\
\textbf{!}Store $count_{\downarrow}/n_{p}$ blocks of vector $q_{old}$ on \\
\textbf{!}$(n_{p}-\textbf{mod}(count_{\downarrow},n_{p}))$ nodes.\\
\indent\indent\textbf{for} $i_{p}=(z+1):n_{p}$\\
\indent\indent\indent\indent\textbf{for} $j=1:r$\\
\indent\indent\indent\indent\indent\indent\textbf{for} $i=1:count_{\uparrow}$\\
\indent\indent\indent\indent\indent\indent\indent\indent$a=a+1$\\
\indent\indent\indent\indent\indent\indent\indent\indent$Q(i,j)=q_{old}(a)$\\
\indent\indent\indent\indent\indent\indent\indent\indent$\lbrace$ Store $Q(i,j)$ on $i_{p}$ node.$\rbrace$\\
\indent\indent\indent\indent\indent\indent\textbf{end}\\
\indent\indent\indent\indent\textbf{end}\\
\indent\indent\textbf{end}\\
\textbf{end}\\
\textbf{!SLAVE PART}\\
\textbf{!}Perform the matrix-matrix multiplication ${\bf{H_{\uparrow}}}Q=vec$.\\
\textbf{if} $\textbf{mod}(count_{\downarrow},n_{p})=0$\\
\indent\indent\textbf{for} $i_{p}=1:n_{p}$\\
\indent\indent\indent\indent$b=(i_{p}-1)r$\\
\indent\indent\indent\indent\textbf{for} $j=1:r$\\
\indent\indent\indent\indent\indent\indent$b=b+1$\\
\indent\indent\indent\indent\indent\indent\textbf{for} $i=1:count_{\uparrow}$\\
\indent\indent\indent\indent\indent\indent\indent\indent\textbf{for} $k=1:count_{\uparrow}$\\
\indent\indent\indent\indent\indent\indent\indent\indent\indent\indent${\bf{H_{\uparrow}}}(k,i)Q(i,j)=vec(k,b)$\\
\indent\indent\indent\indent\indent\indent\indent\indent\textbf{end}\\
\indent\indent\indent\indent\indent\indent\textbf{end}\\
\indent\indent\indent\indent\textbf{end}\\
\indent\indent\textbf{end}\\
\textbf{else}\\
\indent\indent\textbf{for} $i_{p}=1:z$\\
\indent\indent\indent\indent$b=(i_{p}-1)s$\\
\indent\indent\indent\indent\textbf{for} $j=1:s$\\
\indent\indent\indent\indent\indent\indent$b=b+1$\\
\indent\indent\indent\indent\indent\indent\textbf{for} $i=1:count_{\uparrow}$\\
\indent\indent\indent\indent\indent\indent\indent\indent\textbf{for} $k=1:count_{\uparrow}$\\
\indent\indent\indent\indent\indent\indent\indent\indent\indent\indent${\bf{H_{\uparrow}}}(k,i)Q(i,j)=vec(k,b)$\\
\indent\indent\indent\indent\indent\indent\indent\indent\textbf{end}\\
\indent\indent\indent\indent\indent\indent\textbf{end}\\
\indent\indent\indent\indent\textbf{end}\\
\indent\indent\textbf{end}\\
\indent\indent\textbf{for} $i_{p}=(z+1):n_{p}$\\
\indent\indent\indent\indent$b=(i_{p}-1)r$\\
\indent\indent\indent\indent\textbf{for} $j=1:r$\\
\indent\indent\indent\indent\indent\indent$b=b+1$\\
\indent\indent\indent\indent\indent\indent\textbf{for} $i=1:count_{\uparrow}$\\
\indent\indent\indent\indent\indent\indent\indent\indent\textbf{for} $k=1:count_{\uparrow}$\\
\indent\indent\indent\indent\indent\indent\indent\indent\indent\indent${\bf{H_{\uparrow}}}(k,i)Q(i,j)=vec(k,b)$\\
\indent\indent\indent\indent\indent\indent\indent\indent\textbf{end}\\
\indent\indent\indent\indent\indent\indent\textbf{end}\\
\indent\indent\indent\indent\textbf{end}\\
\indent\indent\textbf{end}\\
\textbf{end}\\
\textbf{!MASTER PART}\\
\textbf{!}Collect matrix vec from each node to form\\
\textbf{!}column vector $q_{new}$.\\
\noindent$a=0$\\
\textbf{if} $\textbf{mod}(count_{\downarrow},n_{p})=0$\\
\indent\indent\textbf{for} $i_{p}=1:n_{p}$\\
\indent\indent\indent\indent$b=(i_{p}-1)r$\\
\indent\indent\indent\indent\textbf{for} $j=1:r$\\
\indent\indent\indent\indent\indent\indent$b=b+1$\\
\indent\indent\indent\indent\indent\indent\textbf{for} $i=1:count_{\uparrow}$\\
\indent\indent\indent\indent\indent\indent\indent\indent$a=a+1$\\
\indent\indent\indent\indent\indent\indent\indent\indent$\lbrace$ Collect $vec(i,j)$ from $i_{p}$ node.$\rbrace$\\
\indent\indent\indent\indent\indent\indent\indent\indent$q_{new}(a)=vec(i,b)$\\
\indent\indent\indent\indent\indent\indent\textbf{end}\\
\indent\indent\indent\indent\textbf{end}\\
\indent\indent\textbf{end}\\
\textbf{else}\\
\indent\indent\textbf{for} $i_{p}=1:z$\\
\indent\indent\indent\indent$b=(i_{p}-1)s$\\
\indent\indent\indent\indent\textbf{for} $j=1:s$\\
\indent\indent\indent\indent\indent\indent$b=b+1$\\
\indent\indent\indent\indent\indent\indent\textbf{for} $i=1:count_{\uparrow}$\\
\indent\indent\indent\indent\indent\indent\indent\indent$a=a+1$\\
\indent\indent\indent\indent\indent\indent\indent\indent$\lbrace$ Collect $vec(i,j)$ from $i_{p}$ node.$\rbrace$\\
\indent\indent\indent\indent\indent\indent\indent\indent$q_{new}(a)=vec(i,b)$\\
\indent\indent\indent\indent\indent\indent\textbf{end}\\
\indent\indent\indent\indent\textbf{end}\\
\indent\indent\textbf{end}\\
\indent\indent\textbf{for} $i_{p}=(z+1):n_{p}$\\
\indent\indent\indent\indent$b=(i_{p}-1)r$\\
\indent\indent\indent\indent\textbf{for} $j=1:r$\\
\indent\indent\indent\indent\indent\indent$b=b+1$\\
\indent\indent\indent\indent\indent\indent\textbf{for} $i=1:count_{\uparrow}$\\
\indent\indent\indent\indent\indent\indent\indent\indent$a=a+1$\\
\indent\indent\indent\indent\indent\indent\indent\indent$\lbrace$ Collect $vec(i,j)$ from $i_{p}$ node.$\rbrace$\\
\indent\indent\indent\indent\indent\indent\indent\indent$q_{new}(a)=vec(i,b)$\\
\indent\indent\indent\indent\indent\indent\textbf{end}\\
\indent\indent\indent\indent\textbf{end}\\
\indent\indent\textbf{end}\\
\textbf{end}\\ 

\section{Algorithm III}
\label{algo3}
\noindent$esigno=esign_{trans_{\uparrow}}(per_{\uparrow}(ns),rep_{\uparrow}(ns),trans_{\uparrow}(ns))esign_{oper}$\\
$alls_{i}=0$\\
\textbf{for} $g=1:(per_{\uparrow}(os)-1)$\\
\indent\indent$alls_{i}=alls_{i}+alls(g)*nstate_{\uparrow}(g)$\\
\textbf{end}\\
\textbf{if} $rep_{\uparrow}(os)\neq 1$\\
\indent\indent$alls_{i}=alls_{i}+alls(per_{\uparrow}(os))*(rep_{\uparrow}(os)-1)$\\
\textbf{end}\\
$alls_{j}=0$\\
\textbf{for} $g=1:(per_{\uparrow}(ns)-1)$\\
\indent\indent$alls_{j}=alls_{j}+alls(g)*nstate_{\uparrow}(g)$\\
\textbf{end}\\
\textbf{if} $rep_{\uparrow}(ns)\neq 1$\\
\indent\indent $alls_{j}=alls_{j}+alls(per_{\uparrow}(ns))*(rep_{\uparrow}(ns)-1)$\\
\textbf{end}\\
\textbf{for} $pd=1:perep_{\downarrow}$\\
\indent\indent$pero1=l.c.m.(period_{\uparrow}(per_{\uparrow}(os)),period_{\downarrow}(pd))$\\
\indent\indent$pero2=l.c.m.(period_{\uparrow}(per_{\uparrow}(ns)),period_{\downarrow}(pd))$\\
\indent\indent$hcf1=h.c.f(period_{\uparrow}(per_{\uparrow}(os)),period_{\downarrow}(pd))$\\
\indent\indent$hcf2=h.c.f.(period_{\uparrow}(per_{\uparrow}(ns)),period_{\downarrow}(pd))$\\
\indent\indent$x=pero2/period_{\uparrow}(per_{\uparrow}(ns))$\\
\indent\indent$y=pero2/period_{\downarrow}(pd)$\\
\indent\indent\textbf{for} $i=0:(x-1)$\\
\indent\indent\indent\indent$de=period_{\uparrow}(per_{\uparrow}(ns))*i$\\
\indent\indent\indent\indent\textbf{for} $j=0:y$\\
\indent\indent\indent\indent\indent\indent$di=period_{\downarrow}(pd)*j$\\
\indent\indent\indent\indent\indent\indent$diff=di-de$\\
\indent\indent\indent\indent\indent\indent$trlup(diff)=i$\\
\indent\indent\indent\indent\indent\indent$trldn(diff)=j$\\
\indent\indent\indent\indent\textbf{end}\\
\indent\indent\textbf{end}\\
\indent\indent$normal_{const}=\sqrt{pero1/pero2}$\\
\indent\indent\textbf{for} $ta=0:(hcf1-1)$\\
\indent\indent\indent\indent$alls_{i}=alls_{i}+1$\\
\indent\indent\indent\indent$yy=ta-trans_{\uparrow}(ns)$\\
\indent\indent\indent\indent\textbf{if} $ta<trans_{\uparrow}(ns)$\\
\indent\indent\indent\indent\indent\indent\textbf{while} $yy \textless 0$\\
\indent\indent\indent\indent\indent\indent\indent\indent$yy=yy+period_{\downarrow}(pd)$\\
\indent\indent\indent\indent\indent\indent\textbf{end}\\
\indent\indent\indent\indent\textbf{end}\\
\indent\indent\indent\indent$yyy=yy+1$\\
\indent\indent\indent\indent\textbf{if} $yyy\textgreater hcf2$\\
\indent\indent\indent\indent\indent\indent$zz=\textbf{mod}(yyy,hcf2)$\\
\indent\indent\indent\indent\indent\indent\textbf{if} $zz\neq 0$\\
\indent\indent\indent\indent\indent\indent\indent\indent$yyy=zz$\\
\indent\indent\indent\indent\indent\indent\textbf{else}\\
\indent\indent\indent\indent\indent\indent\indent\indent$yyy=hcf2$\\
\indent\indent\indent\indent\indent\indent\textbf{end}\\
\indent\indent\indent\indent\textbf{end}\\
\indent\indent\indent\indent$diff=trans_{\uparrow}(ns)-ta+yyy-1$\\
\indent\indent\indent\indent$fe=(esignb_{trans_{\uparrow}}(per_{\uparrow}(ns),rep_{\uparrow}(ns)))^{trlup(diff)}$\\
\indent\indent\indent\indent$esigni=esigno*fe$\\
\indent\indent\indent\indent$fi=(esign_{trans_{\downarrow}}(pd,1))^{trldn(diff)}$\\ 
\indent\indent\indent\indent$esign=esigni*fi$\\
\indent\indent\indent\indent$r=alls_{i}$\\
\indent\indent\indent\indent$s=alls_{j}+yyy$\\
\indent\indent\indent\indent$matrix(r,s)=normal_{const}*matel(esign)$\\
\indent\indent\indent\indent\textbf{for} $f=1:(nstate_{\downarrow}(pd)-1)$\\
\indent\indent\indent\indent\indent\indent$fi=(esign_{trans_{\downarrow}}(pd,(f+1)))^{trldn(diff)}$\\
\indent\indent\indent\indent\indent\indent$esign=esigni*fi$\\
\indent\indent\indent\indent\indent\indent$r=r+hcf1$\\
\indent\indent\indent\indent\indent\indent$s=s+hcf2$\\
\indent\indent\indent\indent\indent\indent$matrix(r,s)=normal_{const}*matel(esign)$\\
\indent\indent\indent\indent\textbf{end}\\
\indent\indent\textbf{end}\\
\indent\indent$alls_{i}=alls_{i}+hcf1*(nstate_{\downarrow}(pd)-1)$\\
\indent\indent$alls_{j}=alls_{j}+hcf2*nstate_{\downarrow}(pd)$\\
\textbf{end}\\

\section{Algorithm IV}
\label{algo4}
\noindent$esigno=esign_{trans_{\downarrow}}(per_{\downarrow}(ns),rep_{\downarrow}(ns),trans_{\downarrow}(ns))esign_{oper}$\\
$xx=0$\\
\textbf{for} $pu=1:perep_{\uparrow}$\\
\indent\indent$pero1=l.c.m.(period_{\uparrow}(pu),period_{\downarrow}(per_{\downarrow}(os)))$\\
\indent\indent$pero2=l.c.m.(period_{\uparrow}(pu),period_{\downarrow}(per_{\downarrow}(ns)))$\\
\indent\indent$x=pero2/period_{\downarrow}(per_{\downarrow}(ns))$\\
\indent\indent$y=pero2/period_{\uparrow}(pu)$\\
\indent\indent\textbf{for} $i=0:x$\\
\indent\indent\indent\indent$de=period_{\downarrow}(per_{\downarrow}(ns))*i$\\ 
\indent\indent\indent\indent\textbf{for} $j=0:(y-1)$\\
\indent\indent\indent\indent\indent\indent$di=period_{\uparrow}(pu)*j$\\
\indent\indent\indent\indent\indent\indent$diff=de-di$\\
\indent\indent\indent\indent\indent\indent$trlup(diff)=j$\\
\indent\indent\indent\indent\indent\indent$trldn(diff)=i$\\
\indent\indent\indent\indent\textbf{end}\\
\indent\indent\textbf{end}\\
\indent\indent$normal_{const}=\sqrt{pero1/pero2}$\\
\indent\indent\textbf{if} pu\textgreater1\\
\indent\indent\indent\indent$xx=xx+alls(pu-1)*(nstate_{\uparrow}(pu-1))$\\
\indent\indent\textbf{end}\\
\indent\indent$alls_{i}=0$ \\
\indent\indent\textbf{for} $g=1:(per_{\downarrow}(os)-1)$\\
\indent\indent\indent\indent$hcf=h.c.f.(period_{\uparrow}(pu),period_{\downarrow}(g))$\\
\indent\indent\indent\indent$alls_{i}=alls_{i}+nstate_{\downarrow}(g)*hcf$\\
\indent\indent\textbf{end}\\
\indent\indent$hcf1=h.c.f.(period_{\uparrow}(pu),period_{\downarrow}(per_{\downarrow}(os)))$\\
\indent\indent\textbf{if} $rep_{\downarrow}(os)\neq 1$\\
\indent\indent\indent\indent$alls_{i}=alls_{i}+(rep_{\downarrow}(os)-1)*hcf1$\\
\indent\indent\textbf{end}\\
\indent\indent$alls_{j}=0$\\
\indent\indent\textbf{for} $g=1:(per_{\downarrow}(ns)-1)$\\
\indent\indent\indent\indent$hcf=h.c.f.(period_{\uparrow}(pu),period_{\downarrow}(g))$\\
\indent\indent\indent\indent$alls_{j}=alls_{j}+nstate_{\downarrow}(g)*hcf$\\
\indent\indent\textbf{end}\\ 
\indent\indent$hcf2=h.c.f.(period_{\uparrow}(pu),period_{\downarrow}(per_{\downarrow}(ns)))$\\
\indent\indent\textbf{if} $rep_{\downarrow}(ns)\neq 1$\\
\indent\indent\indent\indent$alls_{j}=alls_{j}+(rep_{\downarrow}(ns)-1)*hcf2$\\
\indent\indent\textbf{end}\\
\indent\indent$yy=trans_{\downarrow}(ns)$\\
\indent\indent$alls_{i}=alls_{i}+xx$\\
\indent\indent$alls_{j}=alls_{j}+xx$\\
\indent\indent\textbf{for} $tt=0:(hcf1-1)$\\
\indent\indent\indent\indent$alls_{i}=alls_{i}+1$\\
\indent\indent\indent\indent$yy=yy+1$\\
\indent\indent\indent\indent\textbf{if} $yy\textgreater period_{\downarrow}(per_{\downarrow}(ns))$\\ 
\indent\indent\indent\indent\indent\indent$fi=yy/period_{\downarrow}(per_{\downarrow}(ns))$\\
\indent\indent\indent\indent\indent\indent\textbf{if} $\textbf{mod}(yy,period_{\downarrow}(per_{\downarrow}(ns))=0$\\
\indent\indent\indent\indent\indent\indent\indent\indent$fi=fi-1$\\
\indent\indent\indent\indent\indent\indent\textbf{end}\\
\indent\indent\indent\indent\textbf{else}\\
\indent\indent\indent\indent\indent\indent$fi=0$\\
\indent\indent\indent\indent\textbf{end}\\ 
\indent\indent\indent\indent$yyy=yy$\\
\indent\indent\indent\indent\textbf{if} $yyy>period_{\downarrow}(per_{\downarrow}(ns))$\\
\indent\indent\indent\indent\indent\indent\textbf{while} $yyy\textgreater period_{\downarrow}(per_{\downarrow}(ns))$\\
\indent\indent\indent\indent\indent\indent\indent\indent$yyy=yyy-period_{\downarrow}(per_{\downarrow}(ns))$\\
\indent\indent\indent\indent\indent\indent\textbf{end}\\
\indent\indent\indent\indent\textbf{end}\\
\indent\indent\indent\indent\textbf{if} $yyy \textgreater hcf2$\\
\indent\indent\indent\indent\indent\indent$zz=\textbf{mod}(yyy,hcf2)$\\
\indent\indent\indent\indent\indent\indent\textbf{if} $zz\neq 0$\\
\indent\indent\indent\indent\indent\indent\indent\indent$xy=zz$\\
\indent\indent\indent\indent\indent\indent\textbf{else}\\
\indent\indent\indent\indent\indent\indent\indent\indent$xy=hcf2$\\
\indent\indent\indent\indent\indent\indent\textbf{end}\\
\indent\indent\indent\indent\textbf{else}\\
\indent\indent\indent\indent\indent\indent$xy=yyy$\\
\indent\indent\indent\indent\textbf{end}\\
\indent\indent\indent\indent$diff=-yyy+xy$\\
\indent\indent\indent\indent$r=alls_{i}$\\
\indent\indent\indent\indent$s=alls_{j}+xy$\\
\indent\indent\indent\indent$fi=fi+trldn(diff)$\\
\indent\indent\indent\indent$esigni=(esignb_{trans_{\downarrow}}(per_{\downarrow}(ns),rep_{\downarrow}(ns)))^{fi}$\\
\indent\indent\indent\indent$esigni=esigni*esigno$\\
\indent\indent\indent\indent$fe=(esignb_{trans_{\uparrow}}(pu,1)))^{trlup(diff)}$\\
\indent\indent\indent\indent$esign=esigni*fe$\\ 
\indent\indent\indent\indent$matrix(r,s)=normal_{const}*matel(esign)$\\
\indent\indent\indent\indent\textbf{for} $f=1:(nstate_{\uparrow}(pu)-1)$\\
\indent\indent\indent\indent\indent\indent$fe=(esignb_{trans_{\uparrow}}(pu,(f+1)))^{trlup(diff)}$\\
\indent\indent\indent\indent\indent\indent$esign=esigni*fe$\\
\indent\indent\indent\indent\indent\indent$r=r+alls(pu)$\\
\indent\indent\indent\indent\indent\indent$s=s+alls(pu)$\\
\indent\indent\indent\indent\indent\indent$matrix(r,s)=normal_{const}*matel(esign)$\\
\indent\indent\indent\indent\textbf{end}\\
\indent\indent\textbf{end}\\
\textbf{end}\\\\

\section{Algorithm V}
\label{algo5}
\noindent$n=0$\\
\textbf{for} $pu=1:perep_{\uparrow}$\\
\indent\indent\textbf{for} $i=1:nstate_{\uparrow}(pu)$\\
\indent\indent\indent\indent$s=0$\\
\indent\indent\indent\indent\textbf{for} $j=1:m$\\
\indent\indent\indent\indent\indent\indent\textbf{if} site $j$ of $repres_{\uparrow}(pu,i)$ is occupied\\ 
\indent\indent\indent\indent\indent\indent\indent\indent$s=s+1$\\
\indent\indent\indent\indent\indent\indent\indent\indent$site_{\uparrow}(s)=j$\\
\indent\indent\indent\indent\indent\indent\textbf{end}\\
\indent\indent\indent\indent\textbf{end}\\
\indent\indent\indent\indent\textbf{for} $pd=1:perep_{\downarrow}$\\
\indent\indent\indent\indent\indent\indent$hcf=h.c.f.(period_{\uparrow}(pu),period_{\downarrow}(pd))$\\
\indent\indent\indent\indent\indent\indent\textbf{for} $k=1:nstate_{\downarrow}(pd)$\\ 
\indent\indent\indent\indent\indent\indent\indent\textbf{for} $j=1:m$\\
\indent\indent\indent\indent\indent\indent\indent\indent\textbf{if} site $j$ of $repres_{\downarrow}(pd,k)$ is occupied\\
\indent\indent\indent\indent\indent\indent\indent\indent\indent\indent$site_{\downarrow}(j)=1$\\
\indent\indent\indent\indent\indent\indent\indent\indent\textbf{else}\\
\indent\indent\indent\indent\indent\indent\indent\indent\indent\indent$site_{\downarrow}(j)=0$\\
\indent\indent\indent\indent\indent\indent\indent\indent\textbf{end}\\
\indent\indent\indent\indent\indent\indent\indent\textbf{end}\\
\indent\indent\indent\indent\indent\indent\indent\textbf{for} $ta=0:(hcf-1)$\\
\indent\indent\indent\indent\indent\indent\indent\indent\indent\indent$p=0$\\
\indent\indent\indent\indent\indent\indent\indent\indent\indent\indent$n=n+1$\\
\indent\indent\indent\indent\indent\indent\indent\indent\indent\indent\textbf{for} $l=1:s$\\
\indent\indent\indent\indent\indent\indent\indent\indent\indent\indent\indent\indent$r=site_{\uparrow}(l)-ta$\\
\indent\indent\indent\indent\indent\indent\indent\indent\indent\indent\indent\indent\textbf{if} $r \textless 1$\\                          
\indent\indent\indent\indent\indent\indent\indent\indent\indent\indent\indent\indent\indent\indent$r=r+m$\\
\indent\indent\indent\indent\indent\indent\indent\indent\indent\indent\indent\indent\textbf{end}\\                
\indent\indent\indent\indent\indent\indent\indent\indent\indent\indent\indent\indent\textbf{if} $site_{\downarrow}(r)=1$\\
\indent\indent\indent\indent\indent\indent\indent\indent\indent\indent\indent\indent\indent\indent$p=p+1$\\  
\indent\indent\indent\indent\indent\indent\indent\indent\indent\indent\indent\indent\textbf{end}\\
\indent\indent\indent\indent\indent\indent\indent\indent\indent\indent\textbf{end}\\
\indent\indent\indent\indent\indent\indent\indent\indent\indent\indent$mat(n,n)=p*U$\\
\indent\indent\indent\indent\indent\indent\indent\textbf{end}\\
\indent\indent\indent\indent\indent\indent\textbf{end}\\
\indent\indent\indent\indent\textbf{end}\\
\indent\indent\textbf{end}\\
\textbf{end}\\

\end{document}